\documentclass[pdflatex,sn-mathphys-num]{sn-jnl}

\usepackage{graphicx}%
\usepackage{multirow}%
\usepackage{amsmath,amssymb,amsfonts}%
\usepackage{amsthm}%
\usepackage{mathrsfs}%
\usepackage[title]{appendix}%
\usepackage{xcolor}%
\usepackage{textcomp}%
\usepackage{manyfoot}%
\usepackage{booktabs}%
\usepackage{algorithm}%
\usepackage{algorithmicx}%
\usepackage{algpseudocode}%
\usepackage{listings}%
\usepackage{lineno}

\usepackage{setspace}

\begin{document}

\title[Article Title]{Pursuit of biomarkers of brain diseases: Beyond cohort comparisons}

\author[1,2,3]{\fnm{Pascal} \sur{Helson}}\email{pascal.helson@u-bordeaux.fr}
\equalcont{These authors contributed equally to this work.}

\author[2,3]{\fnm{Arvind} \sur{Kumar}}\email{arvkumar@kth.se}
\equalcont{These authors contributed equally to this work.}

\affil[1]{\orgdiv{UMR 5293, IMN}, \orgname{University of Bordeaux, CNRS}, \orgaddress{\street{146 Rue Léo Saignat}, \city{Bordeaux}, \postcode{33000}, \country{France}}}

\affil[2]{\orgdiv{School of Electrical Engineering and Computer Science and Digital Futures}, \orgname{KTH Royal Institute of Technology Stockholm}, \orgaddress{\street{Lindstedsvagen 5}, \city{Stockholm}, \postcode{10040}, \country{Sweden}}}

\affil[3]{\orgname{Science For Life Laboratory}, \orgaddress{\street{Tomtebodavagen 23}, \city{Solna}, \postcode{17165}, \country{Sweden}}}

\abstract{
Despite the diversity and volume of brain data acquired and advanced AI-based algorithms to analyze them, brain features are rarely used in clinics for diagnosis and prognosis. Here we argue that the field continues to rely on cohort comparisons to seek biomarkers, despite the well-established degeneracy of brain features. Using a thought experiment (Brain Swap), we show that more data and more powerful algorithms will not be sufficient to identify biomarkers of brain diseases. We argue that instead of comparing patient versus healthy controls using single data type, we should use multimodal (e.g. brain activity, neurotransmitters, neuromodulators, brain imaging) and longitudinal brain data to guide the grouping before defining multidimensional biomarkers for brain diseases.}

\keywords{Brain diseases, Biomarkers, Differential diagnosis, Open data}

\maketitle
\onehalfspacing
\section*{Introduction}
Behavior is the most obvious readout of the brain function. Therefore, the diagnosis of a brain disease entails a number of behavioral assessments (e.g. the Movement Disorders Society-Unified Parkinson's Disease Rating Scale (MDS-UPDRS) for Parkinson's disease (PD)~\cite{goetz2008movement} and the Montreal Cognitive Assessment (MoCA) for Dementia~\cite{nasreddine2005montreal}). Once the behavior-based diagnosis is narrowed down, the process to establish physiological causes or biomarkers of the behavior symptoms can start. Biomarkers are essential for deciding the treatment plan and prognosis of the disease. 

More recently, there is also an impetus to define biomarkers for the diagnosis of brain diseases, especially for early diagnosis. Typically, the biomarkers are searched at the molecular level (e.g. gene expression and chemical imbalances) or morphological level (e.g. loss of neural or glial cells) or brain activity (fMRI/MEG/EEG/ECOG) and network structure (sMRI, PET, connectome) levels. However, despite years of research, such brain-based biomarkers have not superseded the behavior-based markers of brain diseases. It is counter intuitive that the brain does not play a primary role in the diagnosis of its own disorders. It is possible that the current biomarker data is not detailed enough and analysis is not sophisticated enough. 

Advances in AI/ML methods have brought fresh hope to define better biomarkers of brain diseases. At the same time, numerous online repositories now provide different measures of the brain from healthy controls and patients with a specific diagnosis [e.g. OpenNeuro (a large bank of public datasets including different brain diseases, \url{https://openneuro.org/}), Enhancing Neuro Imaging Genetics through Meta-Analysis (ENIGMA, \url{https://enigma.ini.usc.edu/}), the Alzheimer’s Disease Neuroimaging Initiative (ADNI, \url{https://adni.loni.usc.edu/}), the Parkinson’s Progression Markers Initiative (PPMI, \url{https://www.ppmi-info.org/}), the Autism Brain Imaging Data Exchange (ABIDE, \url{https://fcon\_1000.projects.nitrc.org/indi/abide/}), and SchizConnect (a portal for data on schizophrenia, \url{https://schizconnect.org/})] (also see \citep{sandstrom2022recommendations,eickhoff2016sharing}). However, so far AI methods have rarely reached a classification accuracy useful for clinical practice and more crucial, AI models have severe reliability issues \cite{grote2024foundation}. In this comment, we provide possible explanations of why our current approach to search for biomarkers is inherently flawed and then a way forward based on interpretable data analysis.

\section*{The case of brain activity based biomarkers}
The brain is a multiscale system which can be described at many levels spanning from molecules to cells, to circuits, networks and networks of networks \citep{churchland1988perspectives,bell1999levels,hemberger2016comparative}. It stands to reason that "a person's mental activities are entirely due to the behavior of nerve cells, glial cells, and the atoms, ions, and molecules that make them up and influence them" \citep{crick1994astonishing}. This suggests that correlates of animal behavior (function/dysfunction) should be observed in the form of the activity of nerve cells, glial cells and the neurotransmitters/neuromodulators. Among these, neural activity seems to play the most prominent role. Eventually, behavior is expressed through muscle activity which are directly controlled by the spiking activity of neurons. Therefore, changes in the neural activity are essential to manifest any behavior. Consistent with this, every behavior can be decoded from the neural activity of one or more brain regions. Thus, one may infer that all dysfunctions of the brain could be decoded from the brain activity which can form a link between structural/chemical changes and behavior. But then why brain activity is only used as a biomarker in a handful cases such as epilepsy focus detection? We answer this question using an AI model and a thought experiment.

\subsubsection*{Brain Swap}
Neural correlates have been observed in several lab experiments involving simple and repetitive tasks. However, the existence of neural correlates of behavior hides an important piece of information: the same behavior can be elicited by many brain activity patterns and the same neural activity pattern in a given brain region could result in different behaviors given the context and the specific individual \cite{krakauer2017neuroscience}. This is called degeneracy which suggests that many different sets of parameters may lead to the same function \cite{prinz2004similar,degeneracy_2024}. Consequently, the same behavioral symptoms do not imply similar brain activity -- often the opposite is true. For instance, tremor could arise from either basal ganglia or cerebellar circuits dysfunction despite identical clinical presentation \cite{brittain2013many}, similar MDS-UPDRS scores may not imply similar brain activity \cite{helson2023cortex,helson2025non}, tinnitus can stem from auditory cortex reorganization or distributed network dysfunction \cite{langguth2024tinnitus}, hallucinations in schizophrenia emerge via distinct neural pathways across patients \cite{curvcic2017interaction}. Conversely, the same brain activity state may be associated with different behavioral states e.g. beta band oscillations in motor cortical activity are observed both in healthy and patients with Parkinson's disease. Such degeneracy has serious implications for the search of biomarkers of a brain disease.

\begin{figure}[ht!]
\begin{centering}
\includegraphics[width=0.9\textwidth]{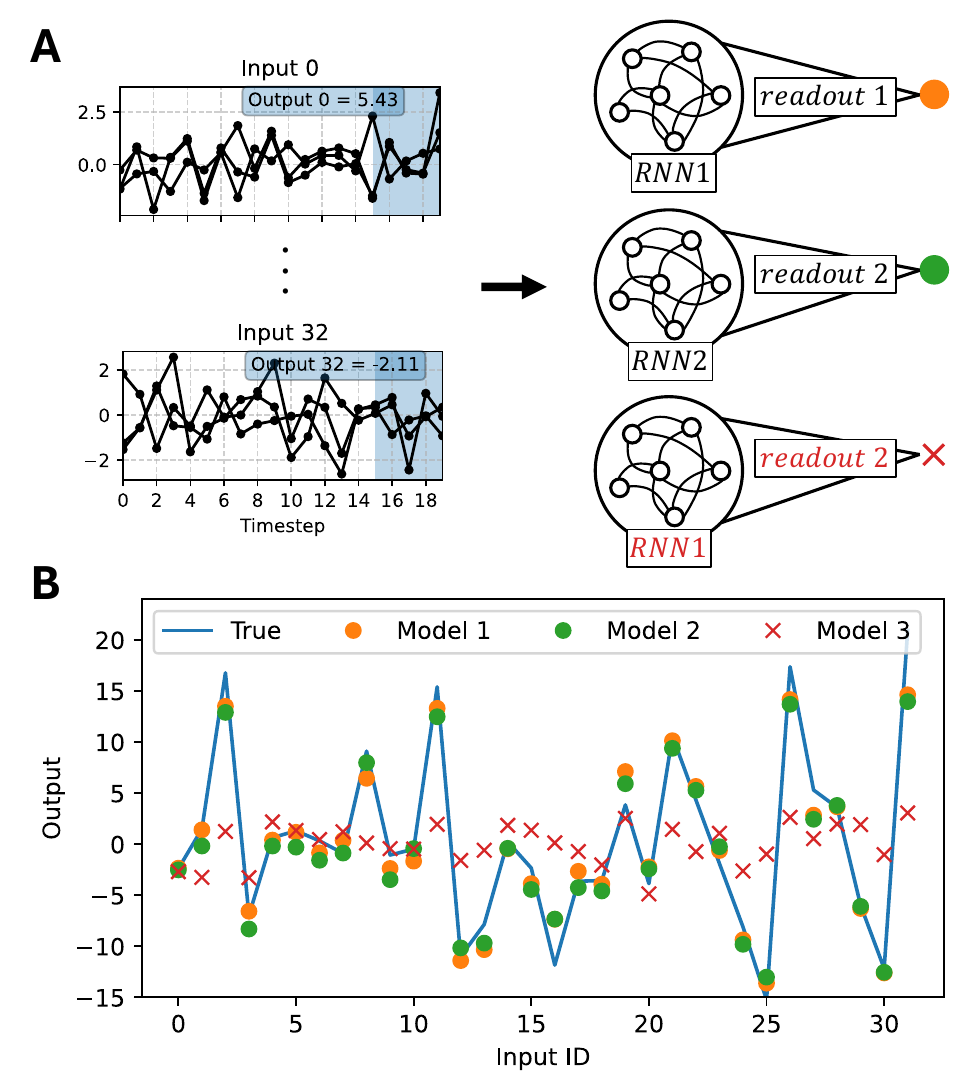}
\par\end{centering}
\caption{\textbf{AI model: learning the same task can be done via different RNN weight optimization.} \textbf{(A)} Left: Example of two time series (inputs) along with their corresponding outputs. Time series are in dimension 3 for easier visualization. The blue area shows the points over which the sum is done to get the output displayed at the top.
Right: Schematic of the models: Model 1 (RNN1 $\to$ readout 1), Model 2 (RNN2 $\to$ readout 2), and Model 3 (RNN1 $\to$ readout 2). We used many-to-one single-layer Elman RNNs with a linear final hidden state readout, mean squared error loss, Adam optimizer, non-linearity $tanh$, hidden layer of size $50$, learning rate $0.01$ and $100$ epochs.}
\textbf{(B)} Comparison of the outputs of the three models versus the true output.
\label{fig:rnn}
\end{figure}

Here, we illustrate this fact using an AI model.  We have deliberately chosen a simple task and neural network -- the claim is that if an idea does not work in a simple setting, it will not work in a more complex biological setting. We considered two RNNs (RNN1 and RNN2) connected to their respective readout layer (Model 1 and Model 2). Both RNNs were identical in their neurons, architecture, and initial synaptic weights (only readouts differed; for more details see the legend of Fig.\,\ref{fig:rnn}A). Both were trained using the same learning rule to perform the same task: predict the sum of the last 5 elements (output) of a time series (input) (Fig.\,\ref{fig:rnn}A). Our input set was composed of 32 time series of time-length 20 and dimension 10 (dimension 3 in Fig.\,\ref{fig:rnn}A), so 32 matrices of size 10$\times$20. Each element of a time series was drawn independently from a normal distribution. Given an input matrix $I=(i_{kt})_{0\leq k \leq 9,0\leq t \leq 19}$, the associated output $O$ was the sum of all the elements of the final 5 time steps, $O=\sum_{t=14}^{19}\sum_{k=0}^{9}i_{kt}$  (Fig.\,\ref{fig:rnn}A). Both RNN1 and RNN2 learned the task to the same proficiency (Fig.\,\ref{fig:rnn}B; loss Model 1: 4.2658, loss Model 2: 5.0760). However, after training, they developed different weights due to the training's stochasticity. That is, the neural activity in both RNNs was 'healthy' -- as it enables the two RNNs to perform the task when connected to their respective readout -- but different.

Next, we changed the setup and connected the output of the first RNN to the second readout (Model 3). As expected the performance of RNN1 degraded significantly (Fig.\,\ref{fig:rnn}B; loss Model 3: 61.2756). Now, based on the task performance, RNN1 appears 'dysfunctional'. We will also get a different dysfunctional response if we instead use the readout layer of the first RNN with the second RNN's activity. 

We can extend this example of RNNs to the brain through a thought experiment: take the motor cortical activity (RNN1) from a healthy person when he/she is performing a reaching task -- the readout here is the mapping from the motor cortex activity (RNN1) to the arm movement (output 1). Map this activity on the motor cortex of another person (replace RNN2 by RNN1). Will the second person also perform the reaching task in the same way as the first person after such a brain swapping? The answer is no. That is, healthy activity of a brain region is 'healthy' given the state of the rest of the brain and would make the other person 'non-functional' or 'dysfunctional'. This thought experiment can be extended to the whole brain network. Indeed, knowing the whole brain activity still excludes downstream decoding of brain activity by spinal circuits and muscles.

The AI model and thought experiment reveal two important issues of biomarker of a brain disease: degeneracy prevents from resorting to cohort comparison and considering single type of data leads to confounding.

\subsubsection*{Cohort comparison approach to identify biomarkers}
When we search for brain activity based biomarkers of a brain disease, we define a certain descriptor of brain activity. Next, we obtain the distribution of this descriptor from a patient group and a matched healthy control group. If we can classify the patients from healthy controls based on the brain activity descriptor, we can argue that we have found a biomarker. This cohort comparison approach implicitly assumes that the brain activity of all healthy subjects (resp. all patients) is statistically identical. But, as we have illustrated with the RNN, thought experiment and degeneracy examples, even identical brain activity could manifest itself as healthy and dysfunctional behavior. That is comparing one cohort against another is not a useful way to identify biomarkers.

Indeed, cases for which comparing cohorts is useful are rare. For example, epileptiform activity in patients with epilepsy or persistent beta oscillations in the basal ganglia of patients with Parkinson's disease are very different from anything seen in healthy subjects. However, in most cases, cohort comparison has not revealed any novel biomarker that can be used in clinical setting \cite{insel2015brain,woo2017building}.

Our example suggests that it is more useful to compare healthy state activity with disease-related activity of the same individual. To paraphrase the line about material models by Rosenblueth and Wiener \cite{rosenblueth1945role}, "a patient is a good model of a patient, preferably the same patient".

\subsubsection*{The problem of subsampling}
In hotel rooms, we have two two-way switches to control the lights in the room -- one at the entrance and the other is typically close to the bed. Can we determine whether the lights are on or off just by looking at the state of one of the two switches? This is not possible because the state of the light bulb is determined by the two switches together. That is, partial sampling is likely to give a random result. In fact, one of the first pseudo-random number generators (middle-square method) invented by von Neumann exactly exploited this fact about partial sampling. 

In our RNN example, the activity and readout weights together allow correct task performance. Just based on the RNN's activity, we cannot predict whether the model will perform the task correctly or not. However, prediction can become possible by incorporating additional context. For instance, if we know that the only difference between two networks is their bias terms -- analogous to neuromodulatory levels like dopamine in the brain -- we can accurately distinguish them and predict their performance. That is, it is crucial to integrate multiple data types to understand neural function and dysfunction.

\begin{figure}[ht!]
\begin{centering}
\includegraphics[width=1.\textwidth]{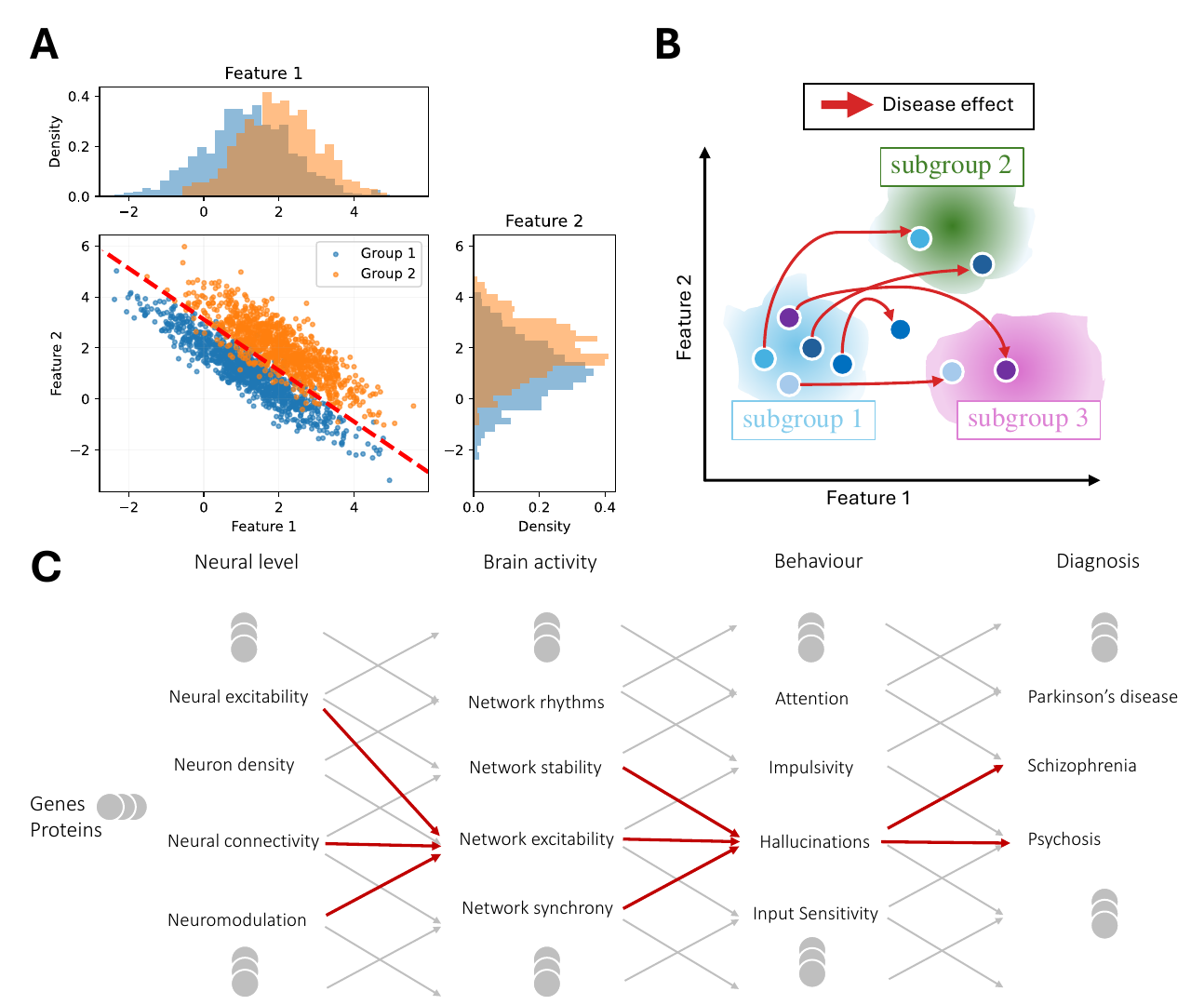}
\par\end{centering}
\caption{\textbf{How multimodal and longitudinal data can help identifying biomarkers of brain diseases.} \textbf{(A)} Simple illustration of data easy to separate when considering two features but not when only one is analyzed. Right and top plots are respectively the marginal distribution of feature 1 and 2 for the two groups.
\textbf{(B)} Assuming you can split your data on healthy controls in three subgroups. When people get a disease, the brain tries to compensate and find another working point -- illustrated by the red arrows -- which may be described by another subgroup's feature collection. If you look at individual changes you will observe some, but if you look at the global level, healthy controls have the same features' distributions than patients.
\textbf{(C)} Example of different possible chains of changes occurring at different levels of the brain leading to the same behavior present in several brain diseases. Same symptoms (e.g. hallucination) could be shared by multiple diseases. The symptom can be manifested in different features of brain activity (red arrow) and those brain activity changes could come from multiple changes at the neuron/synapse level (red arrows). This makes it essential to study and identify correlates of a diseases at multiple scales of brain organization.
}
\label{fig:solution}
\end{figure}

We illustrate this fact in Fig. \ref{fig:solution}A with a simple example showing how data clearly separable when considered in a two dimensional space becomes indistinguishable when sampled from only one dimension. In the brain, the situation is of course more complex and each disorder is affected by a large number of factors. An example of such subsampling is the need of multiple modalities in the detection of both the focal point and the altered network in epileptic subjects \cite{pittau2015functional, fiallo2025multimodal}.

\section*{Beyond brain-activity based biomarkers}
\subsubsection*{The necessity of multimodal data}
The problem in the brain is far more complex than in our AI model and thought experiment. As stated earlier, the brain is a multi-scale system and its function or dysfunction is determined by multiple levels simultaneously (Fig. \ref{fig:solution}C). Sampling only one level or few levels is not likely to give a correct diagnosis. That is, measurements of one of a few of the brain's state variables is bound to fail and outperformed by multimodal analysis; see \cite{suk2014hierarchical, luo2024multimodal} and references therein for examples of Alzheimer’s disease (AD) and mild cognitive impairment (MCI) stages detection.

Indeed, we have focused so far on brain activity but our arguments extend to other potential biomarkers of brain diseases. This is not a surprise that genetic, chemical and morphological features alone are usually not sufficient to classify patients from healthy controls. Counter-examples are rare but there exist for example one gene diseases like Huntington's disease (HD) and the spinal muscular atrophy (SMA) \cite{rust2025elusive}. However, this does not mean that there is a cure: SMA has several -- especially solving this specific gene mutation -- but HD do not have any. On the other hand, several potential biomarkers failed in the past. Indeed, there are examples of 'healthy' subjects with less cortical density than patients with Dementia \cite{stern2012cognitive}.
Similarly, it has been observed that 100s genes are implicated in Schizophrenia but each one may not explain even 1\% of the risk of the disease \cite{schizophrenia2014biological}.

In addition, it would greatly surprise us if a doctor prescribed medicine only using a single biomarker. A major part of medical training of physicians is about differential diagnosis, i.e. after looking at multiple symptoms and clinical indicators, they can identify the most likely causes of the symptoms. For instance, a neurologist observes a subject with a slight tremor in one hand and a shuffling gait. She then assesses these behavioral and motor symptoms with the MDS-UPDRS~\cite{goetz2008movement}. Based on this exam, a clinical diagnosis of Parkinson's disease is made. The core biomarker and cause is the progressive loss of dopamine-producing neurons in the substantia nigra compacta region of the brain. Confirming the dopamine deficit (chemical biomarker) with PET scans directly informs the treatment plan. Consecutive PET scans provide a rate of neuronal loss hence helping the neurologist to provide a prognosis for disease progression.

But this essential part of medicine is completely ignored when it comes to the use of AI to develop/identify novel diagnostic biomarker -- we assume that AI will solve the problem of degeneracy and redundancy by itself \cite{huang2023multimodal}.

The web is full of repositories of various kind of data collected from healthy controls and patients with specific diagnosis. In most cases, these repositories used to contain one specific measurement e.g. levels of protein $\alpha$-synuclein or brain images or brain activity. They now diversified and multimodal as well as longitudinal data are available (e.g. PPMI and ENIGMA).

Based on our arguments and thought experiment, we would claim that a data repository without data about multiple levels of the brain (from genetics to brain activity and behavior) from the same subject is not useful \cite{vieira2024multivariate}.

\subsubsection*{What is a brain disease: Third person versus first person perspective}
The idea of comparing one cohort of healthy controls with a cohort of subjects with a certain disease tacitly assumes that the disease condition pushes the patient out of the healthy distribution. This is a third person perspective on the what should be called a disease. Such longitudinal analysis are already showing promising results for example in epilepsy \cite{barrios2025structural}, AD \cite{fischer2025longitudinal} and PD \cite{waldthaler2026magnetoencephalography} progression.

But there is a first person perspective. An individual compares him/herself against their previous state (longitudinal analysis). If there is a big deviation, then they go for a medical consultation. However, this would mean that every individual tracks their brain activity data and statistically significant deviations with respect to the past data should be taken as signs of a brain disease.

As we have highlighted with our AI model and thought experiment, we need to abandon the cohort comparison approach (third-person perspective) and develop protocols that involve individual specific longitudinal measurements of multimodal (proteins, neurotransmitters, neuromodulator, neural activity etc.) brain data.
Such high dimensional longitudinal data should enable us to subgroup subjects (not patients) \cite{insel2015brain} -- see below the "subgroup hypothesis" -- to gain more understanding on the disease and the possible interactions between the multiple modalities at hand. This is particularly necessary in the case of disorders such as autism spectrum disorders (ASD) \cite{insel2015brain} which have high inter subject variability.

\section*{A way forward}
\subsubsection*{The subgrouping hypothesis}
Despite each brain possessing a unique learning history and a capacity to compensate for slowly accumulating pathologies \cite{degeneracy_2024}, they have evolved to solve similar complex tasks in dynamic environments. This shared functional burden -- the necessity to solve a multitude of survival-critical problems like energy efficiency, perception, action, and learning -- powerfully constrains their evolutionary and developmental possibilities. This principle aligns with the concept of multi-task constraint in systems theory \cite{sterling2015principles, baxter2000model}: the more tasks a system solves, the more constrained it becomes. Just as an artificial neural network trained on multiple tasks must find a shared representation that narrows its viable parameter space, the brain is tuned toward a limited set of robust architectural solutions \cite{sterling2015principles}.

Consequently, the features of a functional brain must lie on rather low-dimensional subspaces (manifolds). These subspaces may consist of several distinct "islands" of solutions (see Fig. \ref{fig:solution}B), but they are vastly fewer in number than the possibilities for a single task (e.g. our AI model). 

Thus, each subject can be described by a dot in a high dimensional space spanned by the features/variables. The "solution island" idea suggests that we should be able to find ways to subgroup subjects within this high dimensional space spanned by multimodal data (Fig. \ref{fig:solution}B). In such a parameter space, longitudinal data is essential to infer how a subject evolves and changes its subgroup membership as disease evolves given specific therapeutic interventions. Thus, patients that evolve along same trajectories could be grouped together.

We refer to this as the subgrouping hypothesis (see Fig. \ref{fig:solution}B). In particular, within a given subgroup, we may observe similar changes due to a disease that we could only observe through longitudinal data. Hence, we must stop looking for a single answer for everyone. Instead, we need to group subjects (not patients) based on their common biological/brain features \cite{insel2015brain} -- the "islands". This is how we will finally run studies that yield clear, statistically powerful biomarkers.

\subsubsection*{Model-driven data analysis}
The subgrouping hypothesis requires identification of appropriate variables which are measured at the correct time point and in sufficient spatio-temporal scales. It is important to note that the variables that define the disease state may not be the ones that we can access using available technology e.g. an EEG recording for a few minutes once a month, or gene expression measured every few months when the subject is in a random state. These constraints clearly make the problem intractable. However, computational models of the brain can help resolve the complexity of this problem. Specifically, computational model of the brain should be developed with following aims in mind:
\begin{itemize}
\item Reveal variables that are otherwise maybe obscured e.g. distribution of synaptic weights, neuronal excitability, excitation-inhibition balance \cite{gao2017inferring}, critical slowing down \cite{scheffer2012anticipating} etc.
Such analyses can be based on mechanistic models -- like The Virtual Brain (TVB) \cite{ritter2013virtual}, spectral graph model (SGM) \cite{raj2020spectral} or Dynamic Causal Modeling (DCM) \cite{friston2003dynamic} -- fitted to the multimodal data. In addition, inference tools -- like Simulation-Based Inference (SBI) \cite{tejero2020sbi} -- allow for more freedom on the model choice and the data statistics to be reproduced such as power spectral density (PSD) \cite{gao2024deep}.
\item Identify variables that are set in a 'sloppy manner' \cite{daniels2008sloppiness} using uncertainty analysis. Such variables will not constitute biomarkers of a brain diseases.
\item Identify variables that bridge scales. Theoretical frameworks from mathematics and physics can define which variables are meaningful and how to relate them across scales. This prevents the "garbage-in-garbage-out" problem of naive data fusion. For example, mean-field theory \cite{wilson1972excitatory, van1996chaos} provides a formal link, showing how the statistics of microscopic neuronal properties (e.g., average firing rates, synaptic strengths) determine macroscopic network states (e.g., asynchronous or synchronized regimes). In addition, one could choose to keep at each scale the parameters which vanish at the next one for example, to avoid redundancy. Similarly, dynamical systems theory, particularly slow-fast analysis \cite{rinzel1998analysis}, offers a concrete strategy. In many neural systems, variables evolve on distinct timescales (e.g., fast spiking vs. slow neuromodulation). This allows for model simplification and suggests that biomarkers should be sought within specific frequency bands or temporal windows that correspond to these separable dynamical processes.
\end{itemize}
These models should be developed in close collaboration with clinicians and experimentalists so as to determine the feasibility to measure theoretically-suggested variables (e.g., LFP beta power as a proxy for network synchrony) and to include practical constraints. This collaboration ensures the analysis remains biologically and clinically relevant.

\subsection*{Interpretable AI}
Modeling is of course not always possible. In such case, AI is then an essential and useful tool to address this huge data analysis problem. Indeed, it is temping to exploit advances such as large language models, and techniques such as mixture of experts, majority voting, which allow for integration of multimodal data for brain disease diagnoses. However, it is not sufficient to just classify but we need to know what was used for classification; known as interpretable AI. Moreover, for clinical applications it is important that explanations are causal.

Recently, the availability of annotated brain connectome provides new frameworks to integrate multimodal brain data \cite{bazinet2023towards} and can reveal the most likely brain regions affected in a disease. In addition, the advances in structure-function knowledge bring an even higher level of interest of graph derived methods \cite{fornito2015connectomics,fotiadis2024structure}. Its combination with AI (e.g. graph neural networks) methods has shown impressive results in disease classification like AD or depression; however, making them interpretable remains an important challenge \cite{bessadok2022graph}.

A combination of model-driven feature engineering can complement AI-based methods to find patterns across massive, diverse datasets that humans simply cannot see \cite{rudin2019stop}. With more data available than ever -- from genetics to brain scans to digital health monitoring -- investing in a model-assisted AI-driven strategy is the most promising path forward for brain health \cite{yang2020artificial,grote2024foundation}. Important breakthrough in the discovery of biomarkers indeed require a collaborative effort between theoreticians, data analysts, experimentalists and clinicians.

\subsubsection*{Summary}
In this \textit{Perspective}, our goal was to expose a crucial problem in our approach to identify biomarkers of brain diseases. We have only provided pointers to potential solutions each of which could be a full article in their own accord. We hope that this perspective will give an impetus to design tools and paradigms to further develop and test the subgrouping hypothesis and thereby improve diagnoses, prognoses and therapies of brain diseases.

\bibliography{sn-bibliography}

\section*{Code availability}
We used Pytorch for our code that you can find \url{https://github.com/paschels/brain_disease_biomarker}.

\section*{Acknowledgments}
We are grateful to Dr. Movitz Lenninger for helpful discussions on the manuscript.

\section*{Authorship}
A.K. and P.H. did all the work together. All authors have read and approved the manuscript.

\section*{Competing interests}
We declare that the authors have no competing interests as defined by Nature Portfolio, or other interests that might be perceived to influence the results and/or discussion reported in this paper.

\section*{Research Funding}
A.K. and P.H. received funding from Digital Futures. A.K. was also partially funded by KTH Strategic Research Initiative Brain Health and StratNeuro program of Swedish research Council.

\end{document}